\documentclass[twocolumn,aps,showpacs,preprintnumbers,amsmath,amssymb,a4paper,floatfix]{revtex4}
\usepackage{graphicx}
\usepackage{amssymb}
\begin{document}
\bibliographystyle{unsrt}
\preprint{}

\title{Accelerated electron populations formed by Langmuir wave-caviton interactions}

\author{N. J. Sircombe}
\email{n.j.sircombe@warwick.ac.uk}
\author{T. D. Arber}%
\affiliation{%
Department of Physics,
University of  Warwick,
Coventry, CV4 7AL,
United Kingdom
}%

\author{R. O. Dendy}
\affiliation{%
UKAEA Culham Division, Culham Science Centre,
Abingdon,
 Oxfordshire, OX14 3DB,
 United Kingdom
 }%

\date{\today}
\begin{abstract}
	Direct numerical simulations of electron dynamics in externally
driven electrostatic waves have been carried out using a relativistic
two-fluid one-dimensional Vlasov-Poisson code.
When the driver wave has sufficiently large amplitude,
ion density holes (cavitons) form. The interaction between these
cavitons and other incoming Langmuir waves gives rise to substantial
local acceleration of groups of electrons, and fine jet-like structures arise
in electron phase space. We show that these jets are caused by
wave-breaking when finite amplitude Langmuir waves experience the
ion density gradient at the leading edge of the holes, and are not
caused by caviton burn-out. 
An analytical two-fluid model gives the critical density gradient and caviton depth 
for which this process can occur. In particular, the density gradient
critically affects the rate at which a Langmuir wave,
moving into the caviton, undergoes Landau damping.
This treatment also enables us to derive analytical estimates for
the maximum energy of accelerated electrons, and
for the energy spectrum along a phase-space jet. These are
confirmed by direct numerical simulations.

\end{abstract}

\maketitle	
\section{Introduction}
\subsection{Outline}
%	The potential of a plasma as a medium for accelerating particles has been known for some time \cite{tajima:1979}. Plasmas can sustain high electric field gradients. In theory capable of accelerating electrons to GeV's over a distance of only a metre. Making tabletop particle accelerators a realistic possibility. Plasma accelerators come in four basic types, laser wakefield acceleration (LWFA), self-modulated LWFA, plasma beat wave accelerators and plasma wakefield accelerators. These basic concepts are detailed in \cite{esarey:1996}. The common factor which unites all these schemes is their reliance on wave-particle resonance to accelerate electrons. 
	This paper describes a novel mechanism for the non-resonant acceleration of electrons within a plasma. We demonstrate how coupling of strong Langmuir turbulence to finite amplitude, coherent Langmuir waves allows the potential energy of the Langmuir waves to be imparted to electrons as the wave breaks. This results in the acceleration of a small population of fast electrons from the background population at the point where wave-breaking occurs. Acceleration occurs spontaneously without requiring an intense laser pulse or particle injection, which are key features of electron acceleration schemes \cite{tajima:1979,bingham:2004,esarey:1996, bulanov:1998,najmudin:2003} that rely on wave-particle resonance, such as laser wakefield acceleration (LWFA), self-modulated LWFA, plasma beat wave accelerators and plasma wakefield accelerators. The electrons accelerated by our new non-resonant mechanism form a collimated phase space jet in the electron distribution function $f_e$. Such populations of energetic electrons are of some concern in inertial confinement fusion (ICF) experiments, because they may contribute directly to capsule pre-heat or form the seed populations for further acceleration. For example, the hot electrons generated by this mechanism could undergo wakefield acceleration \cite{pukhov:2002,umstadter:1996, amiranoff:1998} in the presence of a laser field, creating even more energetic particles capable of pre-heating the ICF capsule \cite{lindl:1995, pesme:2002}.\\
	By considering the relativistic one dimensional Vlasov-Poisson system and exposing this system to a large amplitude electrostatic standing wave (in a similar manner to Ref.\cite{califano:1998}), it is possible to drive the formation of cavitons. These cavitons are, in essence, local regions of ion density depletion accompanied by a local concentration of electric field amplitude. Their formation is driven by the ponderomotive force and they are seeded by small fluctuations in the electron density. In our simulations these initial fluctuations are provided by the  electrostatic driving field. However, once the process of cavitation has begun, the external field can be removed without significantly affecting further caviton development. In this paper we show that at a particular point in the caviton development, fine jet-like structures form in the electron phase space, and we derive the necessary conditions for jet formation depending on the characteristics of the caviton. These phase space jets represent a portion of the electron population accelerated from a small region near the edge of a caviton. The process by which these the jets emerge, together with their characteristics, suggests that they are the result of breaking Langmuir waves. This hypothesis is supported by further simulations, and by a simple model for the process based on a fluid treatment including Landau damping. This theory is an extension of the work of Akhiezer and Polovin \cite{akhiezer:1956} on breaking Langmuir waves in a uniform medium.  The energy distribution and the maximum energy within a jet can be calculated from conservation of energy, provided we assume that all of the potential energy of the wave is imparted to the electrons at the breaking wavefront. 
%The VP System
\subsection{\label{VPsystem}The Relativistic Vlasov-Poisson System}
	The model used in this paper is a one dimensional relativistic Vlasov-Poisson system of electrons and protons with a mass ratio $m_i / m_e = M_r$, and no magnetic field. This fully nonlinear self consistent system is governed by the Vlasov equation for the electron distribution function $f_e (x,p,t)$ 
	\begin{equation}
			\label{vlasov_e}
			\frac {\partial f_e}{ \partial t} + \frac{p}{m_e\gamma} \frac{\partial f_e}{\partial x} - eE\frac{\partial f_e}{\partial p} = 0,
		\end{equation}
	the Vlasov equation for the ion distribution function $f_i (x,p,t)$
		\begin{equation}
			\label{vlasov_i}
			\frac {\partial f_i}{ \partial t} + \frac{p}{m_i\gamma} \frac{\partial f_i}{\partial x} + eE\frac{\partial f_i}{\partial p} = 0,
		\end{equation}
	and Poisson's equation for the electric field
		\begin{equation}
			\label{poisson}
			\frac{\partial E}{\partial x} = -\frac{e}{\epsilon_0} \left( \int f_e dv - \int f_i dv \right)
		\end{equation}
	where $x$ is the spatial co-ordinate, $p = \gamma m_e v$ is the momentum co-ordinate and $\gamma = (1+p^2 / m_e^2 c^2)^{1/2}$ is the Lorentz factor. Taking $\tilde z$ to represent the normalised form of the variable $z$, the following dimensionless normalisation, appropriate to a relativistic system, is adopted throughout: $x = (c  / \omega_{pe})  \tilde{x}$, $t = (1 / \omega_{pe}) \tilde{t}$, $v =  c\tilde{v}$, $E = (\omega_{pe}cm_e / e) \tilde{E}$, $p = m_ec\tilde{p}$.
	It follows that frequencies are normalised to the plasma frequency, $\omega = \omega_{pe} \tilde \omega$, wavenumbers to the ratio of the plasma frequency to the speed of light, $k = \omega_{pe} \tilde k / c$, and temperatures are normalised such that $T_{e,i} = (k_B / {m_{e,i}c^2}) \tilde T_{e,i}$, where $k_B$ is Boltzmann's constant.
	All simulations are carried out using a mass ratio $M_r=100$. This value is sufficiently large to allow the development of phenomena on two disparate time-scales (since electron and ion plasma frequencies are an order of magnitude apart), without the increased runtime of the real ratio.	
\subsection{Numerical Approach}
	The Vlasov-Poisson system is solved using the code detailed in Ref.\cite{arber:2002}. This is a split Eulerian scheme in which the distribution functions ($f_e, f_i$) are calculated on a fixed Eulerian grid, and the solver is split into separate spatial and velocity space updates \cite{cheng:1976}. These updates are one dimensional, constant velocity advections carried out using the piecewise parabolic method \cite{colella:1984}. The original code has been extended to solve the fully relativistic Vlasov-Poisson system Eqs.(\ref{vlasov_e}) to (\ref{poisson}). 
	A large amplitude external driving field $\tilde E_d = \tilde E_0\sin (\tilde k\tilde x) \sin (\tilde \omega_0 \tilde t)$ is added to the self consistent electric field found from Poisson's equation. The parameters of the driving field are chosen to ensure that the system is driven at resonance ($\omega_0 = \omega_{pe}$, or $\tilde \omega_0 = 1$ in normalised units), and that the intensity of the perturbations corresponds to the high quiver velocity regime $v_q^2 / v_{Te}^2 > 1$, where $v_q = e E_0 / m_e \omega_0$. Hence $E_0 > m_e \omega_0 v_{Te} / e$, or in normalised units
	\begin{equation}
		\label{E0_gt_vte}
		\tilde E_0 > \tilde v_{Te}
	\end{equation}
	This driving field, similar to the perturbation used in earlier work on the modulational instability \cite{califano:1998}, is necessary to drive the formation of cavitons and is only needed during the early stages of the simulation. In our simulations it is applied from $\tilde t = 0$ to $\tilde t = 10$, which is sufficient to seed the formation of cavitons. \\
	Extending the presence of the driving field after $\tilde t = 10$ does not significantly effect the evolution of cavitions or the amplitude of Langmuir waves in the system. Hence the energy of accelerated electrons remains $\lesssim$ 5MeV. However, Langmuir waves formed as a result of the the continued presence of $\tilde E_d$ result in more numerous and frequent phase space jets.

%%% RESULTS %%%		
\section{Results}

\subsection{Initial Conditions}
	The system is initialised with a Maxwellian distribution of both electrons and ions at equal temperatures with an electron thermal velocity
	\begin{equation}
		\label{vetherm}
		v_{Te}^2 = {c^2} / {10^3}
	\end{equation}
	and an ion thermal velocity
	\begin{equation}
		\label{vitherm}
		v_{Ti}^2 = {c^2} / {10^3 M_r},
	\end{equation}	
	equivalent to a temperature $\approx 0.5$keV. In normalised units this gives $\tilde v_{Te} = 10^{-3/2}$, hence Eq.(\ref{E0_gt_vte}) implies that we require $\tilde E_0 > 10^{-3/2}$. The simulation box is of length $\tilde L_x = 20\pi$ in normalised units, with periodic boundary conditions in space. 
	In summary, the dimensionless initial conditions are $\tilde{T_e}  =  0.001$, $\tilde{T_i}  =  0.001$, $\tilde{L_x}  =  20 \pi$, $ \tilde{E_0}  = 0.5$, $\tilde{\omega_0}  = 1$, $\tilde{k}  = {2 \pi} / {\tilde{L_x}} $, and the initial maximum of $f_e$, $f_e^{max} = 9.2925$.
\subsection{Caviton Formation and Jet Creation}

	\begin{figure*}
		\begin{center}
		\includegraphics[width=14cm]{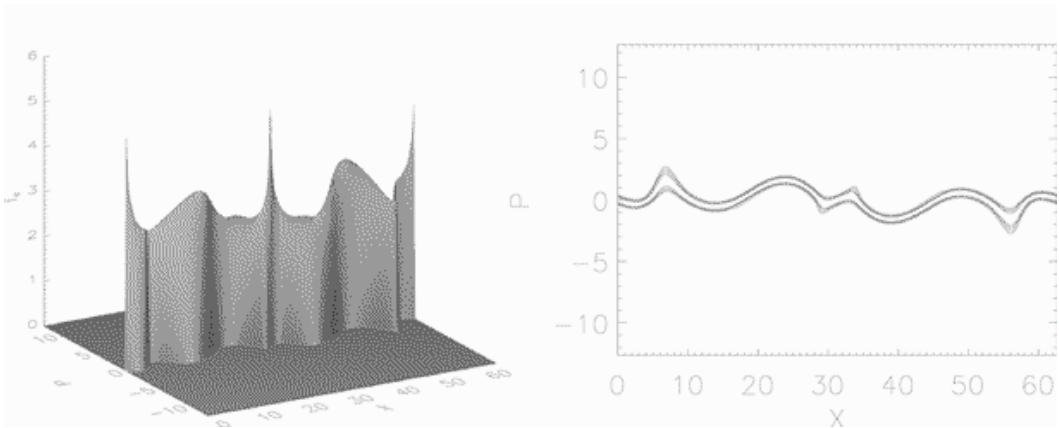}
					\caption{{\em Isometric view of the electron distribution function $f_e$ (\textbf{left}) and contour plot of $\log (f_e > 10^{-6})$ (\textbf{right}) at time $\tilde{t} = 20$. Here the relativistic Vlasov-Poisson system is driven from $\tilde t = 0$ to $10$ at $\omega = \omega_{pe}$. The evacuation of electrons from localised regions of the simulation domain is seen in the electron distribution function; this is the early stage of cavitation. Caviton formation is due to the modulational instability initially driven by the external field $E_d$.}}
					\label{jetFormationT20}
		\end{center}
	\end{figure*}
	\begin{figure*}
		\begin{center}
		\includegraphics[width=14cm]{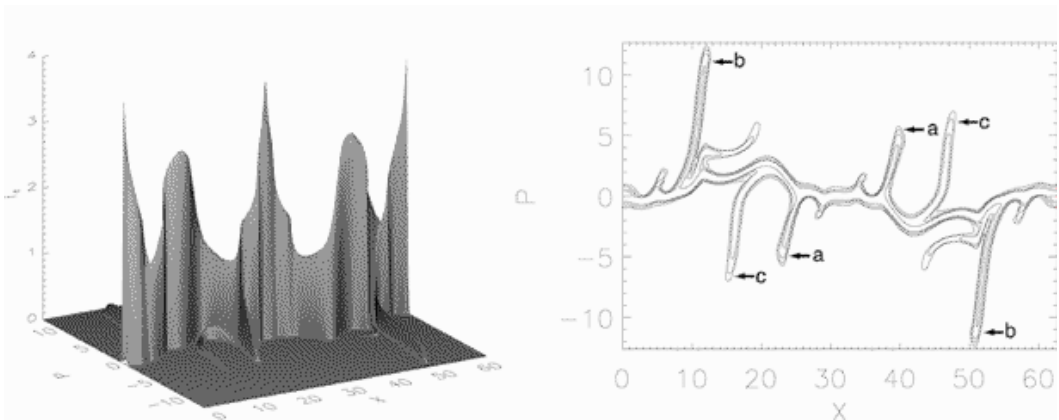}
					\caption{{\em Isometric view of the electron distribution function $f_e$ (\textbf{left}) and contour plot of $\log (f_e > 10^{-6})$ (\textbf{right}) at time $\tilde{t} = 40$. Here the relativistic Vlasov-Poisson system was driven from $\tilde t = 0$ to $10$ at $\omega = \omega_{pe}$. Fine jet-like structures in electron phase space are visible: jets marked \lq\textbf{a}' have recently formed on the inside edges of the two cavitons, whereas the jets marked \lq\textbf{b}' have formed in the same area at an earlier time and have since been advected through the system. The jets marked \lq\textbf{c}' are at an intermediate stage, having formed on the outside edges of the cavitons. These jets are the result of Langmuir wave-breaking at the edges of the evolving density holes.}}
					\label{jetFormationT40}
		\end{center}
	\end{figure*}
	\begin{figure}
		\begin{center}
		\includegraphics[width=0.46\textwidth]{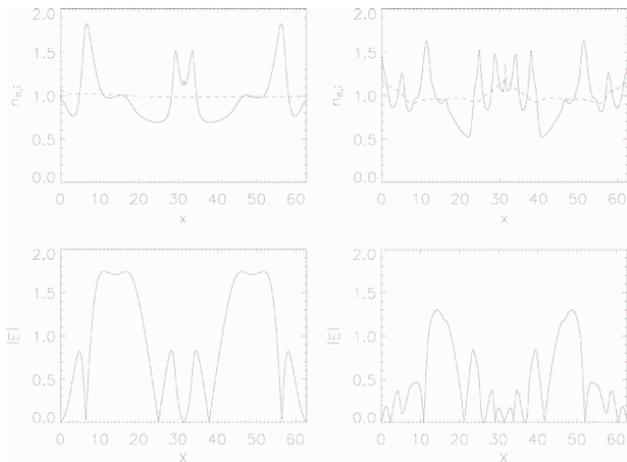}
					\caption{{\em Plots of the electron (\textbf{solid line}) and ion (\textbf{dashed line}) density at $\tilde t = 20$ (\textbf{top left}) and $\tilde t = 40$ (\textbf{top right}), together with the electric field amplitude at  $\tilde t = 20$ (\textbf{bottom left})  and $\tilde t = 40$ (\textbf{bottom right}) for a relativistic Vlasov-Poisson system that was driven from $\tilde t = 0$ to $10$ at $\omega = \omega_{pe}$. The formation of ion density holes can be clearly seen, together with localised growth in the electric field amplitude characteristic of caviton formation. Caviton formation is due to the modulational instability driven initially by the external field $E_d$.}}
				\label{jetFormationN}
		\end{center}
	\end{figure}
	\begin{figure}
		\begin{center}
		\includegraphics[width=0.46\textwidth]{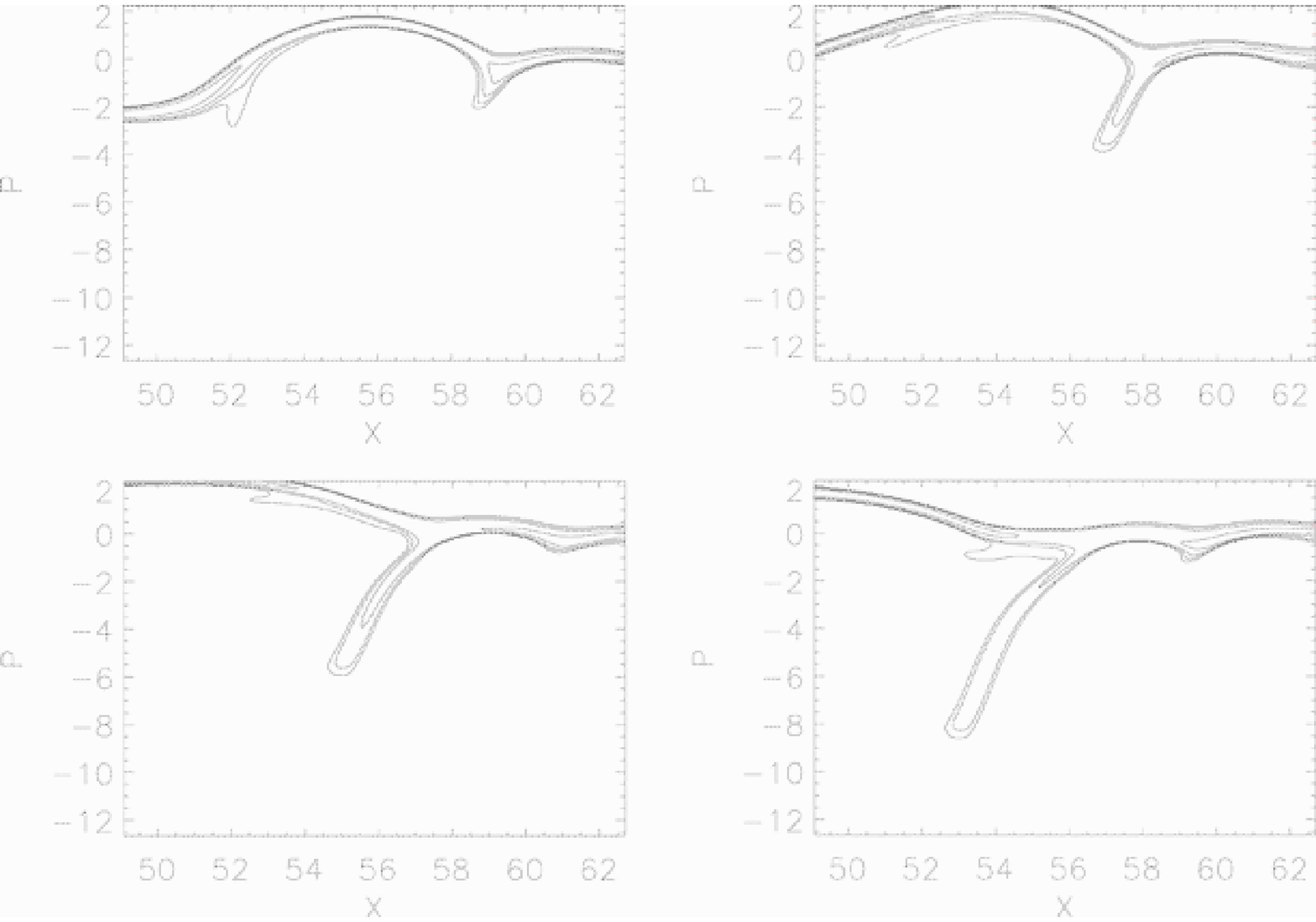}					\caption{{\em Contour plots of $\log (f_e > 10^{-6})$ in the region $50 \lesssim \tilde x \lesssim 60$ at times $\tilde t = 32$ \textbf{(top left)}, $34$ \textbf{(top right)}, $36$ \textbf{(bottom left)} and $38$ \textbf{(bottom right)}. These plots show a phase space jet developing at the outside edge of the caviton (see Fig.\ref{jetFormationN}) at $\tilde x \approx 60$. Electrons are accelerated from a compact region of the background distribution to form the jet, which extends and advects across the caviton.}}
			\label{jetFormationCloseup}
		\end{center}
	\end{figure}
The development of cavitons can be seen in both the isometric renderings of the electron distribution function (Figs.\ref{jetFormationT20} and \ref{jetFormationT40}) and plots of the density and electric field amplitude (Fig.\ref{jetFormationN}). Cavitons form in response to the ponderomotive force exerted on the electrons by the driving field $E_d$. Hence they are centred around regions where the amplitude of the driving field is strongest, leading to symmetry in $n_{e,i}$ and $|E|$ about the centre of the system $\tilde x=10\pi$. By $\tilde t = 20$ two prominent holes, driven by the ponderomotive force, have formed in the electron density centred at $\tilde x \approx 25$ and $37$. As the system continues to evolve, the ions are slowly evacuated to form two broad density holes. These holes in both ion and electron densities are accompanied by a local increase in the electric field amplitude, characteristic of caviton formation, see Fig.\ref{jetFormationN}.\\
At later times, populations of accelerated electrons are clearly visible in the contour plots of the electron distribution function. These are the electron phase space jets. At $\tilde t = 20$, ten plasma periods after the driving field has been removed, there is no evidence of jet formation, see Fig.\ref{jetFormationT20}. However, between $\tilde t = 20$ and $\tilde t = 40$ a series of phase space jets form on both edges of the deepening cavitons. Studying  the evolution of the electron distribution function during this period allows one to identify phase space jets at different points in their evolution. In Fig.\ref{jetFormationT40} we highlight jets at three separate stages of evolution. There are two jets (\textbf{a}) forming on the inside edges of the caviton as well as two old jets (\textbf{b}) which have been advected across the system, effectively crossing the cavitons on whose edges they formed. Finally, there are two intermediate jets (\textbf{c}), on the outer edges of the cavitons. The most energetic electrons within these jets have energies $\approx$ 5MeV. 
	The appearance of these jets does not affect the development of the cavitons, which continue to deepen after the appearance of the phase space jets. In addition, the simulation is seeded with randomly distributed tracer particles. These do not contribute to the numerical solution of the Vlasov Poisson system and are simply moved in response to the total electric field (self consistent electric field plus driving field). Following the motion of these particles indicates that the electrons which form the jets are not accelerated from within the caviton - this effectively rules out caviton burn-out as an explanation for the origin of the phase space jets.
	\subsection{Jet Emergence}
	The process of jet emergence is best explained by observing the evolution of the electron distribution function, focusing on a region where a jet develops, during its early stages. Figure \ref{jetFormationCloseup} shows a reduced section of the electron phase space, the region $50\lesssim \tilde x \lesssim 60$ where the rightmost jet, labelled \lq b' in Fig.\ref{jetFormationT40},  first appears. This region encompasses the right hand side of one of the deepening cavitons  seen in Fig.\ref{jetFormationN}. The jet forms at the outer, right hand edge of this caviton at $\tilde x \approx 60$. The sequence of contour plots in Fig.\ref{jetFormationCloseup} shows that the jet then extends out from the main electron distribution, at $|\tilde p| \approx 2$, to momenta of $|\tilde p| \approx 8$ in approximately one plasma period.
\section{Physics of Electron Jets}
\subsection{Wave Breaking}
\begin{figure}
		\begin{center}
		\includegraphics[width=0.48\textwidth]{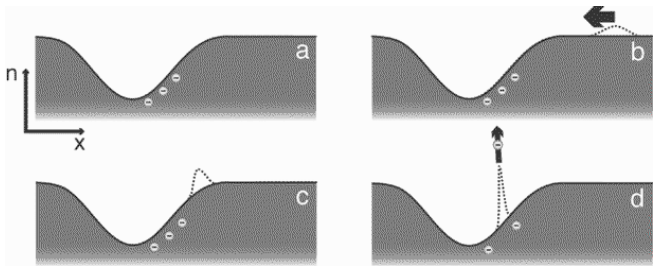}
					\caption{{\em Schematic representation of electron acceleration by Langmuir wave-breaking. \textbf{a:} Density hole with a few sample electrons highlighted.  \textbf{b:} The crest of a Langmuir wave moves towards the density hole.  \textbf{c:} As the wavefront moves down the gradient at the edge of the density hole, its phase velocity falls and the wavefront steepens.  \textbf{d:} For a density hole of sufficient depth and steepness, the incoming wave breaks, forming a strong localised electric field which accelerates electrons from the inside edge of the caviton to create jets in electron phase space.}}
					\label{WaveBreaking}
		\end{center}
	\end{figure}

	The key observed features of the phase space jets are:
	\begin{enumerate}
		\item Jets do not appear at early times in the simulation, they first require some degree of caviton evolution.
		\item Jets are not directly related to the external driver, they appear after the removal of the external driving field.
		\item Jets are not associated with caviton burn-out processes, since cavitons persist long after the appearance of jets.
		\item Jets originate at the caviton edge. Electrons are accelerated from the main distribution to form the phase space jet which is then advected through the system. Their constituent electrons pass over the caviton on whose edge they formed, escaping the influence of the caviton completely. The direction of this advection indicates that phase space jets are the result of processes originating outside the caviton.
		\item Electrons are accelerated up to energies of 5MeV from an initially Maxwellian population with temperature $\approx$ 0.5keV.
	\end{enumerate}
	From extensive numerical simulations of the system, it is clear that the breaking of Langmuir waves on the density gradients at the edges of the cavitons is responsible for the creation of the phase space jets. Figure \ref{WaveBreaking} provides a schematic illustration of the physical process, which we explore in the rest of this section. The Langmuir wave first approaches the density hole. As it moves into the region of lower density, the phase speed at the front of the wave falls. If this proceeds rapidly enough to overcome the effect of Landau damping (which acts to damp the incoming wave energy, and thereby prevent it from breaking), then the wave will break. This creates a strong electric field localised at the wave crest, which accelerates electrons in the vicinity away from the background population to form phase space jets. In the next three sections we outline an analytical model of the competing processes of wave-breaking and Landau damping, together with a derivation of the electron energy distribution within the jet.
	
\subsection{Wave Breaking Amplitude}
	The maximum amplitude $E_{br}$ of a Langmuir wave which can be sustained in a plasma before it breaks is given by the Akhiezer and Polovin constraint \cite{akhiezer:1956}. For a non-relativistic phase velocity $v_p$, this is given by 
		\begin{equation}
			\label{ak_po}
			{eE_{br}} / {m_e\omega_{pe}v_p}=1
		\end{equation}
	where the breaking amplitude $E_{br}$ is the maximum electric field amplitude which can be sustained by a Langmuir wave before it breaks. In normalised units Eq.(\ref{ak_po}) is equivalent to $\tilde E_{br} = \tilde v_p$; the breaking amplitude depends on the phase velocity of the wave, which is determined by the Langmuir dispersion relation
		\begin{equation}
			\label{dispersion}
			{\omega}^2  =  \omega_{pe}^2 + 3 {v_{Te}}^2 k^2 
		\end{equation}
Consider a Langmuir wave moving through a region of decreasing density
		\begin{equation}
			\label{ne}
			\tilde n = \tilde n(\tilde x)
		\end{equation}
	where $n$ falls from an initial value $n_0$, corresponding to a plasma frequency of $\omega_{pe}^0=\left({{n_0 e^2} / {m_e \epsilon_0}}\right)^{1/2}$. As a first approximation, we may assume that the amplitude and frequency of the Langmuir wave remain fixed at $E_L$ ($\tilde E_L$ in normalised units) and $\omega \approx \omega_{pe}^0$ ($\tilde \omega \approx 1$ in normalised units) respectively. Taking the dispersion relation Eq.(\ref{dispersion}) in normalised units
		\begin{equation}
			\label{dispersionNorm}
			{\tilde \omega}^2 = \tilde n(\tilde x)+3 {\tilde {T_e}} {\tilde k}^2
		\end{equation}
		with $\tilde \omega \approx 1$ gives an estimate of the local wavenumber of the Langmuir wave as a function of $\tilde n(x)$:
		\begin{equation}
			\label{estk}
			\tilde k = \left({\left({1 - \tilde n(x)}\right) / {3{\tilde {T_e}}}}\right) ^{1/2}
		\end{equation}	
		This corresponds to a phase velocity $\tilde v_p={\tilde \omega} / {\tilde k} \approx {1}/{\tilde k}$ given by
		\begin{equation}
			\label{estv}
			\tilde v_p = \left({{3{\tilde {T_e}}}/{(1 - \tilde n(\tilde x))}}\right)^{1/2} = \tilde E_{br}
		\end{equation}
		the breaking amplitude, by Eq.(\ref{ak_po}). For the case of a linear density ramp $\tilde n(\tilde x) = 1- \eta \tilde x$ where $0 < \tilde x < 1 / \eta$, Eqs.(\ref{estk}) and (\ref{estv}) imply
		\begin{equation}
			\label{estk_ramp}
			\tilde k = \left({{\eta \tilde x} / {3{\tilde {T_e}}^2}}\right)^{1/2}
		\end{equation}	
		and
		\begin{equation}
			\label{estv_ramp}
			\tilde v_p = \left({{3{\tilde {T_e}}}/{\eta \tilde x}}\right)^{1/2} = \tilde E_{br}
		\end{equation}
	As the wavefront moves down the density ramp, the phase velocity falls, reducing the maximum wave amplitude which can be sustained. If the breaking threshold falls sufficiently, it will be satisfied by the incoming wave, which will then break. The wave breaking condition
		\begin{equation}
			\label{break_cond}
			\tilde E_{br} \leq \tilde E_{L}
		\end{equation}
	together with Eq.(\ref{estk_ramp}) gives
		\begin{equation}
			\label{break_cond2}
			\left({{3{\tilde {T_e}}}/{\eta \tilde x}}\right)^{1/2} \leq \tilde E_L
		\end{equation}
	Thus far, this treatment has not taken into account the effect of Landau damping of the Langmuir wave as it moves down the density ramp, which could reduce the field amplitude at a rate which ensures the breaking condition is never achieved. If the initial amplitude of the Langmuir wave is $\tilde {E_{L0}}$, then the damped amplitude at time $\tilde t$ is given by 
		\begin{equation}
			\label{landau}
			\tilde E_L = \tilde E_{L0} e^{-\Gamma \left(\tilde t\right)}
		\end{equation}
	Here the local damping decrement $\gamma_L$ \cite{krall} determines $\Gamma (t)$		\begin{equation}
			\label{decrement}
			\Gamma \left(\tilde t\right) = \int^{\tilde t}_{0} \gamma_L dt
		\end{equation}
		\begin{equation}
			\gamma_L = {\left(\frac{\pi}{8}\right)}^{\frac{1}{2}} \frac{\omega_{pe}}{k^3\lambda_D^3} \exp{\left({-\frac{3}{2}} {-\frac{1}{2{k}^2 \lambda_D^2} }\right)}
		\end{equation}
	Using the estimate for $\tilde k$ given in Eq.(\ref{estk_ramp}) and solving the integral in Eq.(\ref{decrement}) for $\tilde x$ following the wave, given that
		\begin{equation}
			\label{vg1}
			\frac{\partial \tilde x}{\partial \tilde t} = \tilde v_g = \frac{\partial\tilde\omega}{\partial\tilde k} = \left({3 {\tilde {T_e}} \eta \tilde x}\right)^{1/2}
		\end{equation}
	yields
		\begin{equation}
			\label{landau2}
				\tilde E_L =\tilde E_{L0}  \exp \left(\frac{1}{\eta\sqrt{\tilde T_e}} {\left(\frac{\pi}{2}\right)}^{\frac{1}{2}} \exp \left(-\frac{3}{2}-\frac{3}{2\eta \tilde x}\right) \right) 
		\end{equation}
	Combining Eqs.(\ref{landau2}) and (\ref{break_cond}) gives the condition for wave-breaking, and hence phase-space jet formation
		\begin{eqnarray}
			\label{landauBreak}
			&&\tilde E_{L0}  \exp \left( \frac{1}{\eta\sqrt{\tilde T_e}}{\left(\frac{\pi}{2}\right)}^{\frac{1}{2}} \exp \left(-\frac{3}{2}-\frac{3}{2\eta \tilde x}\right) \right) \nonumber \\
			&&- \left({\frac{3{\tilde {T_e}}}{\eta \tilde x}}\right)^{1/2} \geq 0
		\end{eqnarray}
	for a Langmuir wave with initial electric field amplitude $\tilde E_{L0}$ moving down a linear density ramp with gradient $-\eta$. In the steep gradient limit (i.e. $\eta \rightarrow \infty$), this is equivalent to Eq.(\ref{break_cond2}). To summarise, as the Langmuir wave moves down the density ramp it experiences Landau damping which reduces the wave amplitude from its initial value. The wave-breaking condition is then only satisfied if the decline in density reduces $E_{br}$ to the value where it is equal to the damped Langmuir wave amplitude $\tilde E_L$ given by Eq.(\ref{landau2}). Thus for a density ramp with edge gradient $\eta$ and depth $\Delta$, the wave will break, thereby accelerating electrons to form phase space jets, if Eq.(\ref{landauBreak}) is satisfied for some $\tilde x$, $0 < \tilde x < {\Delta} / {\eta}$. This relation has been derived from a simple fluid treatment, modified to take into account the effect of Landau damping, which depends on caviton gradient $\eta$ and depth $\Delta$. It is supported by numerical simulations using fixed background density profiles containing cavitons with a chosen depth and edge gradient, which are presented below.
	
	\subsection{Fixed Ion Simulations}
\begin{figure}
		\begin{center}
		\includegraphics[width=0.48\textwidth]{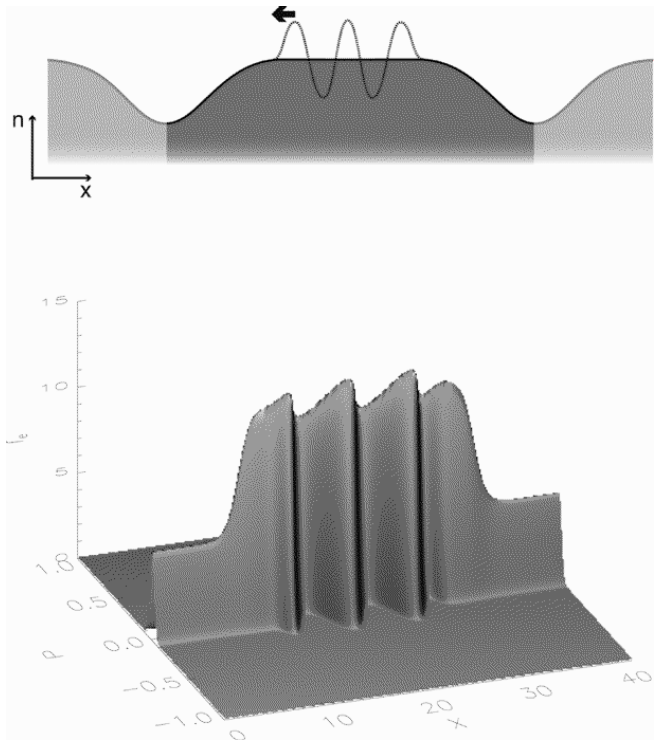}					\caption{{\em \textbf{Top:} Schematic of the initial density profile used for fixed ion simulations. The simulation boundaries are periodic with density holes  defined by a \lq tanh' function. A traveling Langmuir wave of fixed wavelength and amplitude is added in the form of a sinusoidal electron density and velocity perturbation at the centre of the system. \textbf{Bottom:} Isometric rendering of the electron distribution function from the fixed ion simulations at $\tilde{t}=0$.}}
					\label{FixedIon}
		\end{center}
	\end{figure}

\begin{figure}
		\begin{center}
		\includegraphics[width=0.48\textwidth]{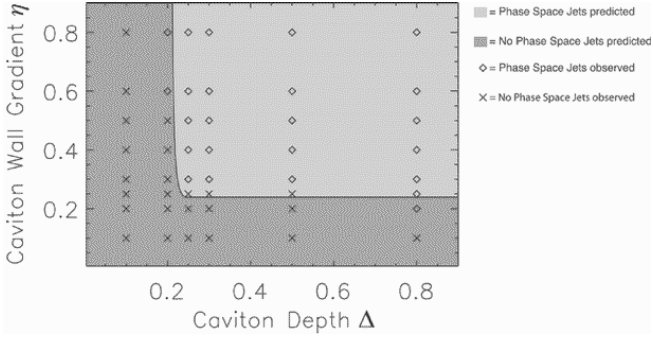}
		\caption{{\em Jet formation and non-formation, plotted with respect to caviton parameter space $(\Delta, \eta)$, where $\Delta$ represents the caviton depth and $\eta$ the edge gradient. The space is partitioned into regions of jet formation and non-formation by the breaking condition Eq.(\ref{landauBreak}), and superimposed are the results from simulations using a fixed ion background to produce the desired caviton structure. The Langmuir wave considered here has amplitude $\tilde E_{L0} = 0.12$ and wavelength $\tilde \lambda_L = 2\pi$.}}
		\label{fixedIonResults}
		\end{center}
	\end{figure}

\begin{figure*}
		\begin{center}
		\includegraphics[width=14cm]{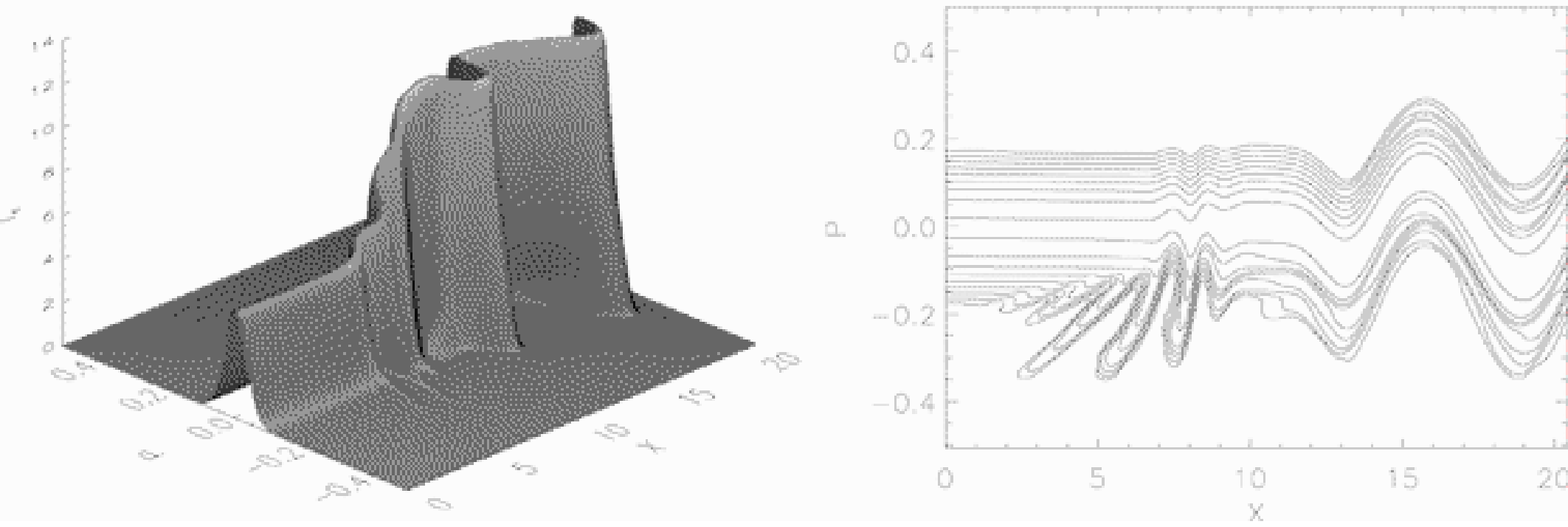}
					\caption{{\em Isometric view of the electron distribution function $f_e$ (\textbf{left}) and contour plot of $\log (f_e > 10^{-6})$ (\textbf{right}) at time $\tilde{t} = 56$ from a simulation of the interaction of a Langmuir wave with an ion density hole (created and maintained via the prescribed immobile ion distribution of Fig.\ref{FixedIon}, top). A Langmuir wave ($\tilde \lambda_L = 2\pi$, $\delta n_e = 0.12$) has encountered the density hole of depth $\Delta = 0.6$ and gradient $\eta = 0.5$. The incoming wave breaks as it moves into the density hole, producing a strong localised electric field which accelerates electrons from the bulk of the distribution to form a jet.}}
					\label{FixedIonT56}
		\end{center}
	\end{figure*}

	In order to test the model outlined above and demonstrate that the phase space jets are a direct result of breaking Langmuir waves, a further series of numerical simulations was conducted. These simulate the interaction of a Langmuir wave with a system having immobile ions whose initial density profile is fixed to provide density holes of a prescribed size and shape, as shown in Fig.\ref{FixedIon}.
A Langmuir wave was created by perturbing the initial electron distribution function to create a traveling wave of a given amplitude and wavelength. This initialisation relies on the linear dispersion relation, and so is only valid for small amplitudes. As the wave propagates across the simulation domain, it encounters the density hole created by the fixed ion background density profile. By varying the parameters of the density hole and the Langmuir wave, it is possible to change the nature of the phase space jets and confirm the functional dependence on caviton depth and gradient predicted by Eq.(\ref{landauBreak}). 
	Figure \ref{fixedIonResults} shows an area of $(\Delta,\eta)$ parameter space partitioned by Eq.(\ref{landauBreak}), for a Langmuir wave of amplitude $\tilde E_{L0} = 0.12$ and wavelength $\tilde \lambda_L = 2\pi$. It also displays the results of simulations of the interaction of this Langmuir wave for a range of $(\eta,\Delta)$, classified by the presence or absence of phase space jets. Our criterion for jet formation is the presence of a second maximum, separated  from the main distribution, in the region of the density ramp. This requires that $f_e$ pass through a critical value $f_c$ four times for some point in the density ramp. The value $f_c$ is chosen to be 1\% of the initial maximum $f_e^{max}$ of $f_e$. The simulation results support the analytical breaking condition Eq.(\ref{landauBreak}) derived above. Figure \ref{FixedIonT56} shows such a simulation at $\tilde t = 56$: a series of phase space jets have formed as a Langmuir wave ($\tilde \lambda_L = 2\pi$, $\delta n_e = 0.12$) encountered a density hole of depth $\Delta = 0.6$ and gradient $\eta = 0.5$.\\
	
\subsection{Electron Energy Distribution within Jets}
\begin{figure}
		\begin{center}
		\includegraphics[width=0.48\textwidth]{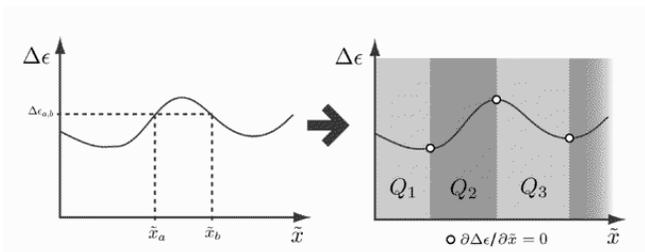}
					\caption{{\em Schematic representation of the decomposition of the spatial domain $S_0$ into sub-domains $\{Q_j \}_{j \in \mathbb{N}}$ at equation (\ref{N2}). The domain $S_0$ represents the region of space from which electrons are accelerated to form a phase space jet. In order to find the energy spectrum, $n(\Delta \epsilon)$ for a phase space jet, we require $\Delta \epsilon (\tilde x)$. However, we must account for the possibility that $\Delta \epsilon$ could be multi-valued with respect to $\tilde x$, for example $\Delta \epsilon (\tilde x_a) = \Delta \epsilon(\tilde x_b) = \Delta \epsilon_{a,b}$. Dividing $S_0$ into  $\{Q_j \}_{j \in \mathbb{N}}$, where the sub-domain boundaries are defined by the condition that ${\partial \Delta \epsilon}/{\partial \tilde x} = 0$, avoids the problem of $\Delta \epsilon (\tilde x)$ being multi-valued. }}
					\label{partition}
		\end{center}
	\end{figure}

\begin{figure}
			\begin{center}
			\includegraphics[width=0.48\textwidth]{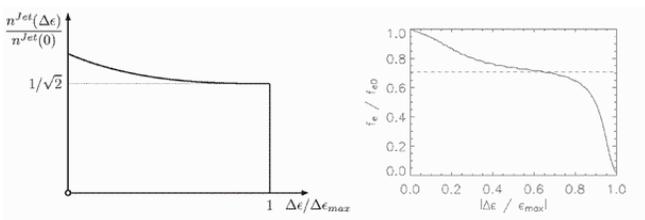}
							\caption{{\em Similarity of predicted and measured forms of the energy distribution of electrons within a jet. \textbf{Left:} Analytical result Eq.(\ref{energy_spec2}). \textbf{Right:} The value of the distribution function $f_e$, normalised to its value $f_{e0}$ at the base of the phase space jet. The energy gain of electrons in the phase space jet is normalised to the maximum observed energy $\approx5MeV$.}}			\label{specs}
			\end{center}
		\end{figure}

	The maximum energy achieved by electrons in a phase space jet, and the distribution of electrons within the jet, can be calculated from the assumption that the total energy carried by the sum of jet electrons comes from the potential of the breaking Langmuir wave. Here we consider the fully relativistic case and show that conservation of energy enables us to find the electron energy distribution along the jet. Following the same approach as Ref.\cite{andreev:1995}, we assume the energy gain of electrons accelerated by the breaking wave is a positive, continuous, single-valued function of their initial position (i.e. their proximity to the breaking wavefront). We then construct an expression for the electron density as a function of energy gain $\Delta \epsilon$. However, we must account for the possibility that two electrons from different spatial positions could achieve the same energy gain. We therefore partition the domain into regions where the number density of accelerated electrons is a single valued function of $\Delta \epsilon$, as shown schematically in Fig.\ref{partition} which we discuss below.
The total number of accelerated electrons $N^{Jet}$ is given by integrating the electron density over the spatial region $S_0$ from which electrons are accelerated:
		\begin{equation}
			\label{N}
			N^{Jet} = \int_{S_0} n(\tilde x_0) d\tilde x_0
		\end{equation}
where $\tilde x_0$ represents the initial position of the electron. This can be rewritten in terms of a new variable $\Delta \epsilon (\tilde x_0)$, the energy gain as a function of initial position, to give
		\begin{equation}
			\label{N2}
			N^{Jet} = \sum_j \int_{Q_j} n_j\left(\tilde x_0\left(\Delta\epsilon\right)\right){\left|\frac{d\Delta\epsilon}{d\tilde x_0}\right|}^{-1}_jd{\Delta \epsilon}
		\end{equation}
Here the region of integration $S_0$ is partitioned into sub-domains $\{ Q_j \}_{j \in \mathbb{N}}$: within each sub-domain $Q_j$, the function $\Delta \epsilon (\tilde x_0)$ is single valued and has no turning points, as illustrated in Fig.\ref{partition}. The boundaries between the sub-domains are defined by the location of the turning points of $\Delta \epsilon (\tilde x_0)$. Denoting the energy spectrum within the jet as a function of energy gain by $n^{Jet}(\Delta \epsilon)$, we can write the total number of electrons in the jet $N^{Jet}$ as an integral of $n^{Jet}$ over $\Delta \epsilon$
		\begin{equation}
			\label{N3}
			N^{Jet} = \int_{S_0} n^{Jet}(\Delta \epsilon) d{\Delta \epsilon}
		\end{equation}
		Equating this to Eq.(\ref{N2}) gives
		\begin{equation}
			\label{energy_spec1}
			n^{Jet}\left(\Delta\epsilon\right) = \sum_j n_j \left(\tilde x_0\left(\Delta\epsilon\right)\right){\left|\frac{d\Delta\epsilon}{d\tilde x_0}\right|}^{-1}_j
		\end{equation}
		The function $x_0(\Delta \epsilon)$  comes from the potential function of the Langmuir wave. Since the electric field of the wave near to breaking will be of  nonlinear sawtooth form, its potential $\tilde \phi_L$ can be represented \cite{andreev:1995} by a parabolic function with amplitude $\tilde\phi_0$
		\begin{equation}
			\label{parabolic}
			\tilde\phi_L = \tilde\phi_0\left(\frac{4{\tilde x_0}^2}{{{\tilde \lambda_L}^2}} - \frac{4\tilde x_0}{\tilde\lambda_L}\right)
		\end{equation}
		over the range of one wavelength $\tilde \lambda_L$. From the assumption that electrons accelerate using the local potential energy of the Langmuir wave, it follows that $\Delta \epsilon(\tilde x_0) = \tilde \phi_L (\tilde x_0)$, with the maximum possible energy gain $\Delta \epsilon_{max} = \tilde \phi_0$: 
		\begin{equation}
			\label{parabolic2}
			\Delta \epsilon = \Delta \epsilon_{max}\left(\frac{4{\tilde x_0}^2}{{{\tilde \lambda_L}^2}} - \frac{4\tilde x_0}{\tilde\lambda_L}\right)
		\end{equation}
		Inverting this expression yields
		\begin{equation}
			\label{invertParabolic2}
			\tilde x_0 = {\tilde \lambda_L}\left(1+{\left(1+{\Delta \epsilon}/{\Delta \epsilon_{max}}\right)}^{1/2}\right)/{2}
		\end{equation}
		while differentiating with respect to $\tilde x_0$ gives
		\begin{equation}
			\label{gradEpsilon}
			\left|\frac{d\Delta \epsilon}{d\tilde x_0}\right| = \Delta \epsilon_{max}\left(\frac{8{\tilde x_0}}{{{\tilde \lambda_L}^2}} - \frac{4}{\tilde\lambda_L}\right)
		\end{equation}
		Substituting Eqs.(\ref{invertParabolic2}) and (\ref{gradEpsilon}) into Eq.(\ref{energy_spec1}), for a constant background density profile, gives the energy spectrum for a single phase space jet: 
		\begin{equation}
			\label{energy_spec2}
			n^{Jet}(\Delta\epsilon) = n^{Jet}(0) \left(\frac{1}{{\left(1+ \Delta\epsilon /\Delta\epsilon_{max}\right)}^{1/2}}\right)
		\end{equation}
		Here we have normalised the equation using $n^{Jet}(\Delta \epsilon = 0)=n_0\tilde \lambda_L / 4 \Delta \epsilon_{max}$, for a local electron density of $n_0$. It follows that the ratio of highest energy to lowest energy electrons within the jet is given by
		\begin{equation}
			\label{ratio}
		 	{n^{Jet}\left(\Delta\epsilon_{max}\right)}/{n^{Jet}\left(\Delta\epsilon = 0\right)} = {1}/{\sqrt{2}}
		\end{equation}
		The analytically derived energy spectrum Eq.(\ref{energy_spec2}), shown in Fig.\ref{specs} (left), is broadly consistent with the electron distribution function that is obtained from the simulations discussed earlier. Figure \ref{specs} (right) shows a plot of $f_e$, normalised to its value $f_{e0}$ at the base of the phase space jet, along the ridge of the phase space jet shown in Fig.\ref{jetFormationCloseup}.

\section{Conclusions}
	Shortly after exposing a fully relativistic Vlasov-Poisson system to a strong external driving field at resonance $\omega_0 = \omega_{pe}$, we observe the formation of fine structures in the electron distribution function, corresponding to the acceleration of small populations of electrons to high energies ($\approx 5$MeV). These phase space jets result from the breaking of Langmuir waves (initially excited by the driving field) in the density gradients of cavitons formed via the ponderomotive force exerted by the external driving field. Restricting the presence of the driving field to $\tilde t < 10$ allows direct acceleration by the driving filed to be ruled out as a mechanism.
	We are able to explain the origin of the small populations of electrons which form the jets; derive a necessary condition for the formation of  jets starting from a basic fluid treatment, which is supported by further numerical simulations; and estimate the energy distribution of electrons within the phase space jet.
	This process may arise whenever Langmuir waves move through a strong density gradient, and is not limited to one dimension or to caviton formation. It may therefore require consideration in laser-plasma interaction contexts spanning inertial confinement fusion and particle acceleration.

\begin{acknowledgments}
This work was supported in part by the United Kingdom Engineering and Physical Sciences Research Council.
\end{acknowledgments}
\bibliography{../Bibliography/refs}
 \end{document}